\documentclass[twocolumn,aps,amsmath,amssymb,showpacs]{revtex4}

\usepackage{graphicx} 
\usepackage{epsfig}

\begin{document}

\title{Unraveling the fluctuations of animal motor activity}   
\author{C. Anteneodo$^{a}$ and D. R. Chialvo$^{b}$   
}
\address{  (a)  Departamento de F\'{\i}sica, PUC-Rio and  
National Institute of Science and Technology for Complex Systems,    
 Rua Marqu\^es de S\~ao Vicente 225, CEP 22453-900 RJ, Rio de Janeiro, Brazil \\
 (b)  Department of Physiology, Feinberg Medical School,
Northwestern University, 303 East Chicago Ave. Chicago, IL 60611, USA  
 }

\begin{abstract}

Human motor activities are known to exhibit scale-free long-term correlated fluctuations  
over a wide range of timescales, from few to thousands of seconds.   
The fundamental processes originating  such fractal-like behavior are
not yet understood. 
To untangle the most significant features of these fluctuations, in this work 
the problem is oversimplified by studying a much simpler system:
the spontaneous motion of rodents, recorded during several days.
The analysis of the animal motion reveals a robust scaling
comparable  with the results previously reported in humans. It is shown that 
the most relevant features of the experimental results can be replicated  by 
the statistics of the activation-threshold model proposed in another context 
by  Davidsen and Schuster.    
\end{abstract}


\maketitle

\section{Introduction}

The timing between consecutive human actions, or even spacing the most simple 
and inconsequential motor activities~\cite{hu2004,nakamura,amaral},  
is known to be scale-free, and despite recent efforts~\cite{barabasi}, 
their mechanisms remain poorly understood. 
The interest is further underscored by  
its potential usefulness as statistical markers for the 
detection and follow up of human neurological disorders. 
The lack of plausible models to account  for these statistical features  calls for 
alternatives able to identify the essential underlying mechanisms.
To that end,  it may be advantageous to uncouple the interference of memory and other 
cognitive factors, present in the experiments with human subjects,  
by analyzing a more elementary process, namely the records of long-term 
spontaneous motor activity of laboratory animals. 

In the present work, the activity of rats was recorded during several days.
and  the experimental data series analyzed from the perspective of a point process.  
The results show striking similarities with the scaling demonstrated to be exhibited
by humans in more elaborated settings. The paper is organized as follows:
In the next section the experimental details are described. Section III contains the statistical
analysis, first for the inter-event times, then for the rates of motion events and finally for the
variance of counts. Section IV describes an activation-threshold model which is able to 
replicate the most relevant experimental observations. Finally, section V closes the paper with 
a  discussion of further implications of the present results.

\section{Experimental setting and data recording}

Male 4-month-old Wistar rats were kept in a soundproof room temperature 
($\sim 20^o$C$-22^o$C) and humidity ($\sim  50-80$ percent) conditioned chambers.
Animals were individually isolated into transparent cages of 25cm$\times$25cm$\times$12cm 
with food pellets and tap water ad libitum. 
They were exposed to a 24-hour cycle of light-dark conditions:
cycles of 12 hs of light (fluorescent lamps with intensity of 300 lux) and 12 hs 
of darkness (dim red light with intensity less than 0.1 lux)~\cite{lab1}.

%
\begin{figure}[b!]
\centering 
\includegraphics*[bb=45 195 560 600, width=0.40\textwidth]{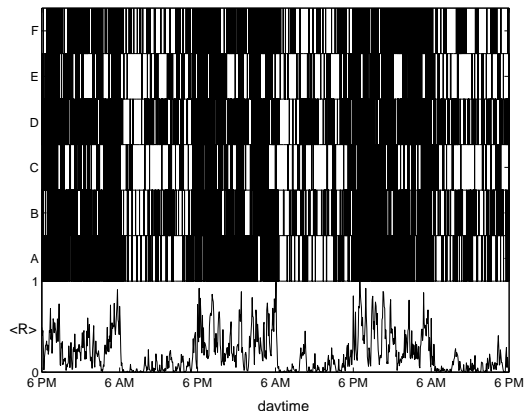}  
 \caption{ Three days activity plot obtained
 from six laboratory animals (``A" trough ``F") exposed to a cycle of 12 hours of
 light (6AM-6PM) and darkness (6PM-6AM), individually
 housed and continuously monitored by an infrared device scanning
 the field at a rate of 1 Hz. The bottom timeseries,  
 depicts the group average activity
 $\langle R \rangle$, computed with a binwidth of 1 min and normalized.
}
\label{fig:tseries}
\end{figure} 
 
Animal's movement was monitored with infrared activity-meters, 
able to scan each rodent housing cage and to report, at a frequency of 1 Hz, animal 
movements  even in absence of locomotion.  Given the sensor's sensitivity
the detection includes even minute head motion, grooming, etc.
An example of the recording for the first three days is plotted in Fig.~\ref{fig:tseries}, 
where changes from activity to immobility, and viceversa, 
are marked by a vertical bar, with a resolution of 1~s. 
In the bottom plot of Fig.~\ref{fig:tseries}, we also show the group average of 
the activity rate $R$, which is a coarse graining of the raw data,
exhibiting the well known circadian rhythmicity.

Our interest here focuses exclusively on understanding the dynamics 
of the irregular fluctuations  and not on the circadian periodicity.

\section{Statistical properties}

\begin{figure}[h!]
\centering 
\includegraphics*[bb=100 500 500 750, width=0.5\textwidth]{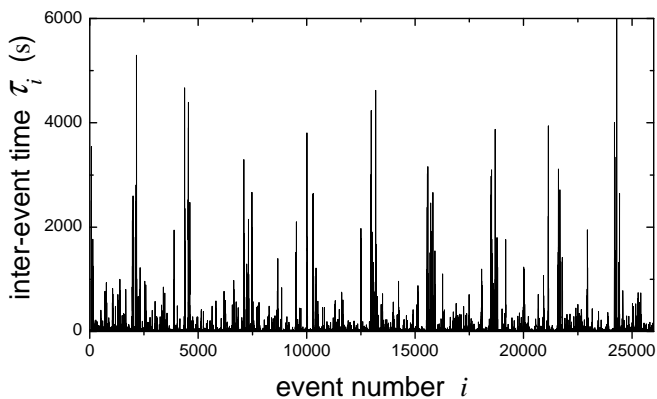}   
\caption{ Inter-event time $\tau_i$ as a function of the event number $i$. (Data from animal A.)}
\label{fig:itimes}
\end{figure}

The spatial and temporal resolution of our experimental setting is such that, 
from the original (binary) data  recorded, we can precisely estimate the duration of
motion (sequences of 1's) and immobility (sequences of 0's) episodes. Thus, the animal 
motor activity can be interpreted as a point process where the beginning of a motion event 
occurs at a definite discrete time $t_i$. 
The point process can be then specified  by the sequence of event times $t_i$ or,  
alternatively, by the series of increments (or inter-event intervals) $\tau_i=t_{i+1}-t_i$.  
A typical plot of the $\tau$-series is presented in Fig.~\ref{fig:itimes}. 
It is evident that, beyond the circadian rhythmicity, clustering or bursts of activity occur.

In order to characterize this point process we will first consider the  
$\tau$-series, obtaining its distribution and correlation structure. Next, 
rates of events will be inspected. Finally,  the statistics of event's 
counts~\cite{thurner} as a function of observation times will be described.
 
\subsection{Inter-event intervals}

\begin{figure}[h!]
\centering 
\includegraphics*[bb=90 370 520 680, width=0.5\textwidth]{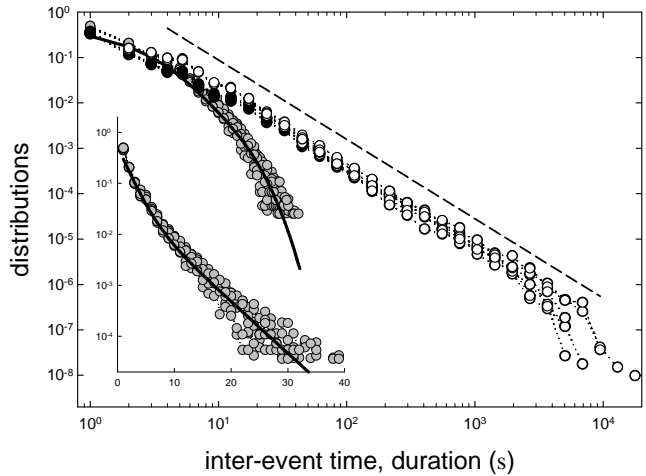}  
 \caption{ 
Normalized distributions of:  inter-event times $\tau$ (empty symbols),
 duration of motion episodes (gray symbols) and duration   of 
 immobility periods (black symbols), computed from nine days of continuous recording. 
Symbols joined by dotted lines  correspond to the results from each one of the six animals. 
The solid lines correspond to a double exponential fit 
(with characteristic times of the order of 1 s and 4 s).
The dashed line, drawn for comparison, has slope -1.75.
Inset: representation of the distribution of motion episodes in log-linear scale. 
Statistics computed individually for each of the six animals and plotted overimposed}
\label{fig:itimes_pdf}
\end{figure}

Fig.~\ref{fig:itimes_pdf} shows the estimation of the probability density $P(\tau)$ 
through the relative frequency of occurrence 
of inter-event times. It was  computed over the nine-day activity (binary) recordings, 
for each one of the six animals. 
We have overimposed the plots from  all the animals to demonstrate the robustness of the results.
It can be clearly seen that, from few seconds to several thousands seconds (about 1 hour),  
the distribution of inter-event times decays as a power-law (with exponent $\mu$ falling  
within the interval $1.75\pm0.5$ for all six animals). 

In contrast, the distribution of duration of motion episodes do possess a 
characteristic timescale (see also Fig.~\ref{fig:itimes_pdf}). 
It can be described by a superposition of two  exponentials with characteristic times 
of the order of 1 s and 4 s, close to the smaller data resolution and to the average duration of 
motion episodes, respectively. 
Let us note that, for human (arm) motor activity data, instead of two exponentials, 
a stretched exponential fit was reported~\cite{nakamura}.

Meanwhile, quiescence intervals are also power-law distributed  
with exponent  $\mu$. 
Because motion intervals are in average much shorter than quiet ones, 
then, the statistics of inter-event times is mainly dominated by that of 
immobility periods, in particular, sharing the same power-law decay. 
Hence, discrepancies between 
both histograms are evident for small time intervals only (see Fig.~\ref{fig:itimes_pdf}).
For comparison, let us note that,  for the power-law exponent of the 
cumulative histogram of inactivity periods of human (arm) motor activity,  
the values $0.92$ (control) and $0.74$ (depressed individuals), 
with about 10\% relative error, have been reported~\cite{nakamura}.  

Identical distributions to those exhibited in 
Fig.~\ref{fig:itimes_pdf} were obtained when day and night 
fluctuations were analyzed separately, only differing in the cutoffs, occurring 
at longer activity (inactivity) intervals, at day (night). 
Therefore the fundamental statistical features are
independent of the level of activity.

\begin{figure}[h!]
\includegraphics*[bb=175 350 510 570, width=0.5\textwidth]{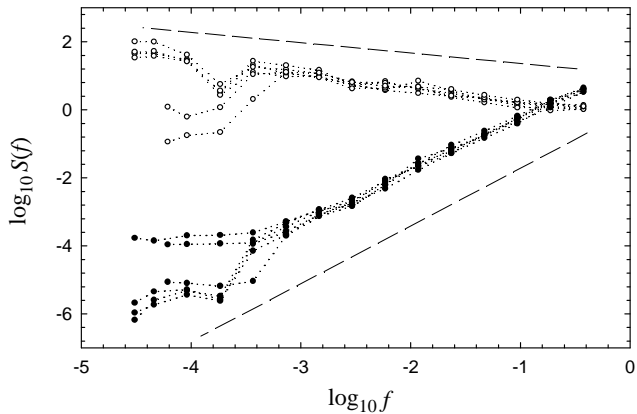} 
\caption{Long-range correlations of inter-event times. 
Log-log plot of the power spectra for the six series of inter-event intervals (open symbols) and 
of their increments (filled symbols). Data were logarithmically binned. 
Here $f$ is in units of the inverse event number. Dashed lines are a guide to the eyes, with slopes $-0.3$ and $1.7$.
Notice that, $\alpha=0.3$ and $\beta=1.7$, verify $\alpha=2-\beta$. 
(Data from six animals.)}
\label{fig:itimes_correl}
\end{figure}

To assess correlations, first we computed the spectral density of  $\tau$,  displayed in 
Fig.~\ref{fig:itimes_correl},   
showing that inter-event intervals are not independent. 
Therefore, the stochastic point process of motor activity can not be considered a renewal one. 
Because of the relatively small value of the exponent of the  $\tau$  power spectrum, 
 we computed also that of the  increments, which helps to confirm that it is not white noise. 
Although not shown, we verified also that motion and quiescent intervals are anticorrelated. 

\subsection{Local rates}
 
\begin{figure}[h!]
\centering 
\includegraphics*[bb=90 110 540 700, width=0.5\textwidth]{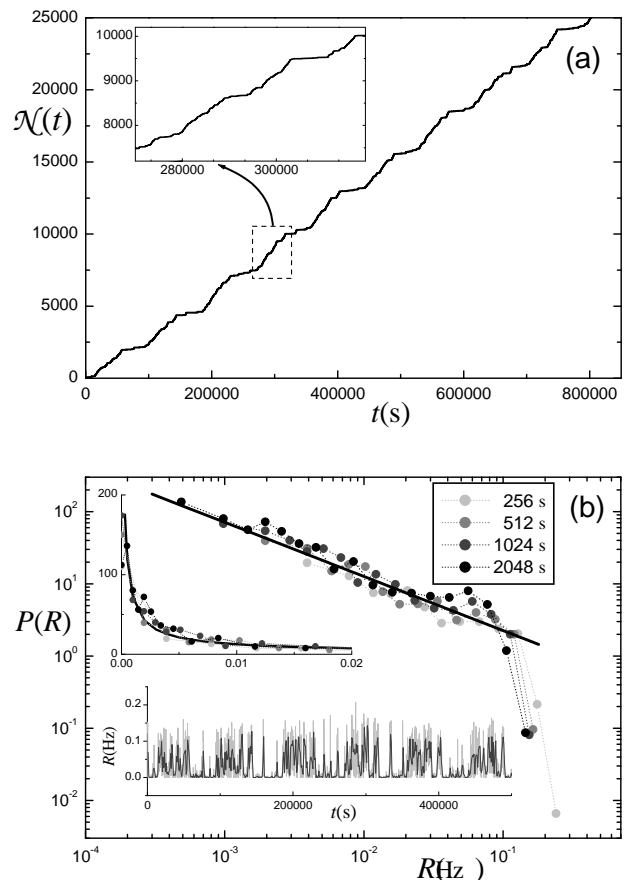}   
 \caption{ (a) Cumulative  number of events as a function of time. A zoom is displayed in the inset.  
 (b) Distribution of local rates $R$ computed over non-overlapping time windows of lengths  $T$
 indicated on the figure. 
 The solid line corresponds to a power-law with exponent $\gamma = 0.75$, 
 drawn for comparison.
 Upper inset: the same plot in linear scale.
 Lower inset: timeseries of local rates $R$ for two of the time window lengths considered.  
 (Data from animal A.) }
\label{fig:rates}
\end{figure}

By dividing the whole observation time interval in $W$ 
(nonoverlapping) uniform windows of length $T$ and counting the number of events, 
$N_n$, in each time window $n$, one obtains the series  of counts.  
The cumulative number of events, ${\cal N}(t)=\sum _{n=1}^mN_n$ vs. time $t=\sum_{n=1}^m \tau_n$,   
is shown in Fig.~\ref{fig:rates}(a). 
One observes that the local rate (local slope) is not constant as in standard Poisson processes. 
It varies on time, following not only the  slowly circadian trend 
but also other rapidly changing ones.

In order to depict more precisely rate inhomogeneities, we 
estimated local rates $R_n$, as being the ratio of $N_n/T$. 
In the scale of half-day, one observes two characteristic mean rates  associated to the 
half-periods of low and high activity levels. 
However, shorter time windows $T$ reveal a more complex structure of the distribution of rates, 
that is not simply bimodal but has the shape displayed in Fig~\ref{fig:rates}(b). 
It follows as a power-law (with exponent $\gamma\simeq 0.75$) with exponential cut-off. 
The higher probability of small rates leads to longer intervals of 
immobility. 
Notice also that there is a range of window lengths for which the distribution of rates 
remains invariant, a scaling feature typical of a fractal-like process. 
Besides this scaling property, rates fluctuate in time, as exhibited in the lower inset of 
Fig.~\ref{fig:itimes_pdf}(b). 
Nonhomogeneous Poisson processes with  (uncorrelated) stochastic rates 
have been considered to explain the emergence of scaling in the statistics of 
inter-event times (see for instance~\cite{hidalgo}). Since in the present case the  inter-event intervals 
are not independent,  such inhomogeneous Poisson processes can be excluded as responsible 
for this dynamics.
 
\begin{figure}[h!]
\includegraphics*[bb=160 325 500 550, width=0.5\textwidth]{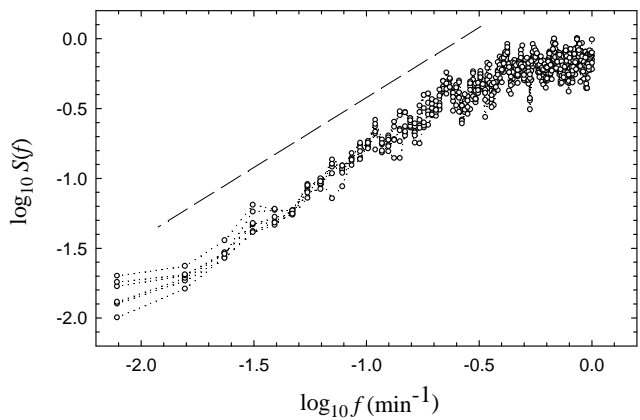} 
\caption{Long-range anticorrelations in the increments $I_n$ of activity rates   
 $R_n$  computed over 1-min windows. 
Log-log plot of the power spectra for the six $I$ timeseries (symbols).
The dashed line with slope $\beta=1$ indicates the correlations expected 
for pink noise. (Data from six animals).
}
\label{fig:rate_correl}
\end{figure}

To estimate rate linear  correlations,  we performed a spectral analysis of 
the time series of increments $I_n=R_n-R_{n-1}$. 
Fig.~\ref{fig:rate_correl} is a log-log plot of its power spectrum $S(f)$.   
It is approximately linear on a wide range of
biologically relevant temporal scales implying that $S(f) \sim f^{\beta}$, 
with $\beta \sim 1$. Since $\beta > 0$, consecutive values of the process $I$ are negatively correlated, 
meaning that increases in activity are, on the average, more likely to be followed 
by decreases and viceversa. 
Shuffling the timeseries of increments yields white noise. 
Recalling that $R$ is the integration of $I$, 
then the spectral density of the original $R$ timeseries decays   
as $1/f^{\alpha}$ with $\alpha=2-\beta$~\cite{Peng93}.

\subsection{Variance of counts}

\begin{figure}[t!]
\centering 
\includegraphics*[bb=130 120 480 340, width=0.5\textwidth]{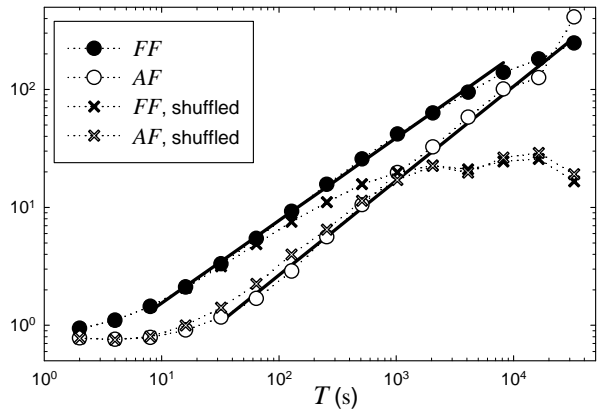}    
 \caption{
Fano and Allan factors as a function of the 
counting time-window length $T$. Solid lines are the results of fits giving 
slopes $d\simeq 0.7$ and $d\simeq 0.8$, respectively.
For comparison, the same analysis over the series of shuffled time intervals is also 
shown. (Data from animal A.)
 }
\label{fig:fano}
\end{figure}

Common measures for detecting correlations in sequences of counts are the 
Fano ($FF$) and Allan ($AF$) factors~\cite{thurner}. The former is the ratio of the variance to the 
mean of the number of events in each time-window, 
$FF=(  \langle [N_k]^2\rangle-\langle N_k\rangle^2)/\langle N_k\rangle$,
an index of dispersion of counts. 
The latter  quantifies  the discrepancy of counts between 
consecutive windows, being $AF=(  \langle [N_{k+1}-N_k]^2\rangle)/(2\langle N_k\rangle)$.  
Although related, they  reflect different features, therefore 
we kept track of both of them. 
In Fig.~\ref{fig:fano},  the two factors are plotted as a function of the length of the 
counting time window $T$. 
There is a range where they increase as $T^d$. The exponent of $FF$ is  $d \in[0.65,0.73]$ 
(that is $d\simeq\mu-1$, within error bars), 
for the six laboratory animals, while that  of $AF$ is about 0.1 higher. 
The discrepancy may be due to the fact that $FF(AF)$ tends to the power law from above(below).  
Their power-law behaviors  indicate that 
the temporal support of the spikes of activity is a fractal-like set with dimension $d$.
Shuffling the series of inter-event intervals modifies both factors, mostly by reducing 
the upper bound of the power-law scaling region. 
This indicates that the scaling properties are partially due to the distribution 
of inter-event intervals itself, but also that
 some correlations are associated with the specific ordering 
of the intervals.  
 

%

\section{Modeling the dynamics}
 
The observed super-Poissonian behavior  
discussed in the previous Section points in the 
direction of a multiplicative or 
clustering process, where the occurrence 
of an event increases the probability of a subsequent one~\cite{teich}. 
A class of  clustering Poisson process was introduced by Gr\"uneis~\cite{gruneis90},  
where there is a primary Poisson process triggering the occurrence 
of a sequences of events (clusters), each following a secondary Poisson process. 
In each cluster the number $m$ of events is a random variable. 

The statistical properties of this two-stage cascade 
are determined also by the cluster size distribution $p(m)$. 
A special case of interest in the context of $1/f^\alpha$ fluctuations is 
$p(m)\propto 1/m^\eta$, for $m\le N_0$, and null otherwise. 
For $3/2<\eta<3$, in the limit $N_0\to\infty$, 
it was reported~\cite{gruneis90} that  the exponent of the variance/mean curve is 
$d \simeq (7-2\eta)/4$ while the exponent of the spectral density 
is $3-\eta$. 
If $\eta=2$, $1/f$ noise is obtained and $d = 3/4$, which is in good accord with  
the present outcomes. 
Although Gr\"uneis et al.  clustering Poisson process can reproduce 
some of the observed scalings, it is not clear how such a cascade process 
would precisely be originated in the present context.
  
The same difficulty applies to other fractal or fractal-rate stochastic point 
processes~\cite{thurner}. 
Recall also that, for many processes cited in Ref.~\cite{thurner}, 
solely the distribution of rates is scale-free, while inter-event times present a 
more trivial statistics. 
In the present case, however, inter-event times are not only scale-free but also present 
long-range correlations, indicating non renewal processes. 

\begin{figure}[t!]
\centering 
\includegraphics*[bb=165 125 550 660, width=0.5\textwidth]{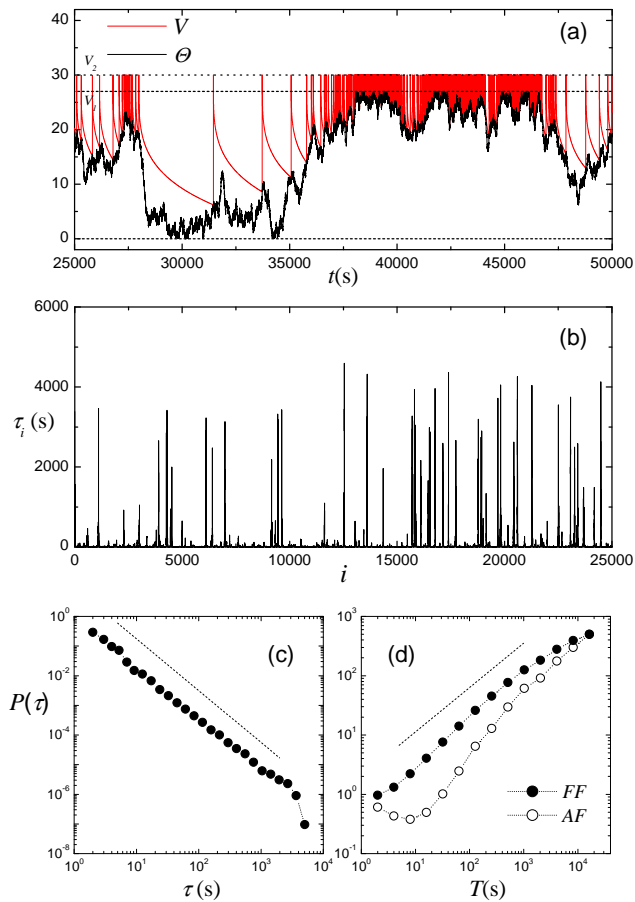}    
 \caption{
 Relaxation toward a fluctuating threshold model. 
 (a) Plots as a function of time of: the threshold $\Theta$ (black line), 
 that fluctuates diffusively (with coefficient $D$) between the levels 
 0 and $V_1$ (dashed lines); the dynamical variable $V$ (thin red line) that 
 describes the relaxation from the excited level $V_2$ (dotted line) down to the threshold 
 and is characterized by amplitude $K$ and exponent $\kappa$. 
 (b)-(d) Results of simulations performed up to time $10^6$ in units that correspond approx. to 1 s of 
 real time. Model parameters are $(V_{1},V_{2},D,K,\kappa)=(27, 30, 0.04, 3.0, 0.25)$.
 (b) Series of inter-event intervals. 
 (c) Distribution of inter-event times. Dotted line indicates $\mu=1.75$.
 (d) Fano and Allan factors vs. counting window length $T$.  Dotted line corresponds to $d=0.7$.   }
\label{fig:model}
\end{figure}

It is plausible that animal activity  is triggered when an internal dynamical state variable  
reaches some value. Without being too specific, an animal can move to eat when sugar level
reaches some low value, for instance. Of course, biological reality will indicate that  
nothing in the judgement of the state variable nor in the threshold value can be very precise. 
Therefore, one can imagine a quantity relaxing towards a fluctuating threshold that resets upon crossing it.
Variants of this scenario are very common in the the literature where 
fluctuations are introduced in either the threshold level or in the 
activation function~\cite{wing,schoner,wagenmakers,th_model}. 
In particular, we examine here  the variant introduced by 
Davidsen and Schuster~\cite{th_model}, 
where the threshold fluctuates following a (bounded) Wiener process  with 
diffusion constant $D$. 
In Fig.~\ref{fig:model}(a), we present a representative example of the fluctuating threshold $\Theta$ and 
the relaxation dynamical variable $V$ as functions of time.  

%

Fig.~\ref{fig:model}(b)-(d) show the model dynamics. 
The distribution of inter-event times depends on the specific shape of the 
decay of $V(t)$, the relaxation process. 
If the decay has the shape $V(t)=V_2-K(t-t_{last})^\kappa$, where $t_{last}$ is the time 
of the previous adjacent trigger,  then 
the power-law distribution of inter-event times has exponent $\mu =2-\kappa$~\cite{th_model}. 
In particular $\kappa=0.25$, yields $\mu=1.75$ close to the observed values, 
and as soon as $\kappa$ approaches zero (abrupt decay), the exponent can increase 
up to a value slightly smaller than 2.  
This relaxation ruled by $\kappa$ might explain the observation in Ref.~\cite{nakamura} 
of $\mu\sim 1.7$ in depressed individuals and $\mu\sim 1.9$ in control ones.  

Besides the inter-event time distributions, the other main  correlation features seen in the data 
are reproduced as displayed by the $FF$ and $AF$ measures (notice, though, the 
relatively larger dip for short $T$ in the $AF$).
It is known that the model dynamics yields $1/f$ spectral properties~\cite{th_model}.
We have verified (not shown) that the inter-event times are power-law correlated and that the
distribution of rates is scale free, as observed for real data,  
although the exponents are different for the chosen set of parameter values. 

Overall, this simple model is able to reproduce the observed phenomenology and it is 
known to be robust under the  addition of noise over the activation signal~\cite{th_model}. Finally,  
the duration of individual events could also be incorporated easily into the model
by an additional integration process consistent with the observed statistics.

\section{Final observations}  

In summary, the results show that animal motion is scale free 
in all of the relevant statistics analyzed and that the
scaling is introduced by the length of the
inactivity pauses as well as its specific ordering.
 
These results are robust for all animals studied, and invariant when day and night data are 
analyzed separately. While motion episodes do possess a definite timescale and 
are basically exponentially distributed, the distribution of inter-event intervals 
are scale-free over  many timescales, from a few  up to thousands of seconds.  
Scaling is also demonstrated in the the rates of occurrence, rejecting a non-homogeneus
Poisson process as responsible for the dynamics.

At even longer time scales than the ones considered here,  many human actions 
exhibit also bursty patterns, in which   
short time intervals of intensive activity are (frequently) separated
by long gaps without  activity~\cite{barabasi}.
In that case, the power-law tails of inter-event distributions have 
been  attributed to queuing processes (where the queue represents the tasks on a priority list).
Depending on the dynamics of the priority list, different classes of universality 
can emerge~\cite{barabasi}. Probably this is not the case in the present experiments, 
where more likely, an  elementary physiological  and/or environmental  mechanism  is behind. 

Some of the scale-free properties here observed, have been reported before 
for human motor activity data as well~\cite{hu2004,amaral}. 
The quantitative aspects for the scaling laws were reported to remain 
unchanged under usual daily activities or periodic scheduled work. 
These observations  were
consistent with earlier results from the study 
of timeseries of heartbeat intervals~\cite{Peng93,Meesmann1,Meesmann2} from healthy humans. 
Then, it have prompted the suggestion of multi-scale physiological
mechanisms as responsible for the presence of long-term correlated dynamics~\cite{hu2004}. 
Other related approaches~\cite{bak1} consider that animal motion could be visualized 
as the output of a large nonlinear dynamical system (e.g., the brain-body-environment 
ensemble) whose repertoire includes the kind of dynamics observed in these experiments. 
Finally, the possibility that the complexity of the environment in itself
influences animal behavior needs to be carefully considered~\cite{foraging}.

The  specific mechanism discussed here, involving a fluctuating-threshold 
controlled dynamics is  biologically plausible, and not only can replicate observed features of the 
statistical structure of rat motor activity, but with  appropriate modifications can be used to 
provide insights about the long-term rhythms alterations 
observed on individuals with mood disorders, depression 
and other neurological disorders, such as chronic pain. 
\\[5mm] \noindent
{\bf Acknowledgements:} DRC acknowledges support by NIH NINDS of USA (Grants NS58661).  
CA acknowledges Northwestern University for the kind hospitality and 
Brazilian agencies CNPq and Faperj for partial financial support.

\end{document}